\documentclass[aps,prc,twocolumn]{revtex4}
\usepackage{latexsym} % Gets \Box etc
\usepackage{amssymb}  % \gtrsim, \geqslant, etc etc: see amsguide.ps
\usepackage{amsmath} % Screws up \beq and \beql! Why?
\usepackage{graphicx}       % For PostScript figures
\usepackage{color}
\usepackage{bm}   % Bold math: \bm{1}

\definecolor{Red}{rgb}{1.00, 0.00, 0.00}

\definecolor{Green}{rgb}{0.00, 0.75, 0.00}

\definecolor{Blue}{rgb}{0.00, 0.00, 1.00}

\definecolor{Cyan}{rgb}{0.00, 0.80, 0.80}

\definecolor{Magenta}{rgb}{1.00, 0.00, 1.00}

\definecolor{Yellow}{rgb}{0.90, 0.90, 0.00}

\definecolor{DarkRed}{rgb}{0.8, 0.0, 0.0}

\definecolor{DarkGreen}{rgb}{0.00, 0.3, 0.00}
\definecolor{DarkBlue}{rgb}{0.00, 0.00, 0.7}
  
\definecolor{DarkCyan}{rgb}{0.00, 0.60, 0.70}

\definecolor{DarkMagenta}{rgb}{0.70, 0.00, 0.70}

\definecolor{DarkYellow}{rgb}{0.9, 0.8, 0.0}

\definecolor{Gold}{rgb}{1.0, 0.84, 0.00} % same as xfig Gold

\definecolor{MediumBlue}{rgb}{0.0, 0.3, 1.00} % not used
 
\definecolor{MediumGreen}{rgb}{0.50, 1.00, 0.50}

\definecolor{PaleBlue}{rgb}{0.8, 0.9, 1.00}
 
\definecolor{PaleGreen}{rgb}{0.7, 1.00, 0.7}

\definecolor{PaleRed}{rgb}{1.00, 0.8, 0.8}

\definecolor{PaleYellow}{rgb}{1.0, 1.0, 0.5}

\definecolor{PaleMagenta}{rgb}{1.00, 0.8, 1.00}

\definecolor{PaleCyan}{rgb}{0.6, 1.0, 1.0}

\definecolor{PaleGrey}{rgb}{0.9, 0.9, 0.9}

\definecolor{Orange}{rgb}{1.00, 0.5, 0.00}

\definecolor{DeepPink}{rgb}{1.00, 0.1, 0.6}

\definecolor{Violet}{rgb}{0.5, 0.00, 0.6}

\definecolor{Brown}{rgb}{0.54, 0.27, 0.07}

\newcommand{\la}{\lambda}

\newcommand{\beq}{\begin{equation}}
\newcommand{\eeq}{\end{equation}}
\newcommand{\ba}{\begin{array}}
\newcommand{\ea}{\end{array}}
\newcommand{\bea}{\begin{eqnarray}}
\newcommand{\eea}{\end{eqnarray}}
\newcommand{\bi}{\begin{itemize}}  %\setlength{\itemsep}{0\parsep}}
\newcommand{\ei}{\end{itemize}}
\newcommand{\ben}{\begin{enumerate}} %\setlength{\itemsep}{0\parsep}}
\newcommand{\een}{\end{enumerate}}
\newcommand{\bc}{\begin{center}}
\newcommand{\ec}{\end{center}}
\newcommand{\bt}{\begin{tabular}}
\newcommand{\et}{\end{tabular}}

\newcommand{\dsp}{\displaystyle}
\newcommand\eqn[1]{(\ref{#1})}      % parentheses around the LaTex "ref" macro
\newenvironment{tightlist}[2]{ 
\begin{list}{#2}{
  \usecounter{#1}
  \setlength{\topsep}{0ex} % added to \parskip
  \setlength{\itemsep}{-\parsep} % added to \parsep
  \settowidth{\labelwidth}{#2} % width of item label
  \setlength{\labelsep}{0.2em}        % space between label and text
  \setlength{\leftmargin}{\labelwidth}% margin = width + separation
  \addtolength{\leftmargin}{\labelsep}
 }}{\end{list}}
\begin{document}

\title{Ghostly action at a distance: \\
a non-technical explanation of
the Bell inequality}

\author{Mark G. Alford}

\affiliation{Physics Department, Washington University,
Saint~Louis, MO~63130, USA}

\begin{abstract}
We give a simple non-mathematical explanation of Bell's inequality.
Using the inequality, we show how the results of Einstein-Podolsky-Rosen
(EPR) experiments violate the principle of strong locality,
also known as local causality. This indicates,
given some reasonable-sounding assumptions,
that {\em some} sort of faster-than-light influence is present in nature. 
We discuss the
implications, emphasizing the relationship between EPR and the Principle
of Relativity, the distinction between causal influences and signals, and
the tension between EPR and determinism.
\end{abstract}

\date{16 Jan 2016} % Hardwire date so arXiv does not change it

% \preprint{}

\maketitle

\section{Introduction}
\label{sec:intro}

The recent announcement of a ``loophole-free'' observation of
violation of the Bell inequality \cite{Hensen:2015ccp} has brought 
renewed attention to the 
Einstein-Podolsky-Rosen (EPR) family of experiments in which
such violation is observed.
The violation of the Bell inequality is often described
as falsifying the combination of ``locality'' and ``realism''.
However, we will follow the approach of
other authors including Bell
\cite{Bell:1976,Maudlin:2011,Norsen:2015}
who emphasize that the EPR results violate a single principle,
strong locality.

Strong locality, also known as ``local causality'',
states that the probability of an event depends only
on things in the event's past light cone. Once those
have been taken into account the event's probability
is not affected by additional information about things
that happened outside its past light cone.

%the probability of an event does not depend
%on things that happened outside its past light cone.
Given some reasonable-sounding assumptions about causation
(see Sec.~\ref{sec:uncertainty}), the violation of  strong locality 
in EPR experiments implies that 
there are causal influences that travel faster than light.
The main goal of this paper is to give an extremely simple non-technical
explanation of how EPR experiments lead to this striking conclusion.
We do this by mapping the experiment onto a situation where
twins are separated and then asked questions to test whether they
can influence each other via faster-than-light telepathy.

Since the EPR results force us to accept that
nature does not respect strong locality, it is natural to
ask how the results cohere with other locality
principles. Is there a sense in which the results violate
``local realism''? 
We discuss Einstein's Principle of Relativity (Lorentz
invariance) and
the principle that signals cannot travel faster than light
(``signal locality'').
We describe how signal locality arises from the Principle of Relativity, 
and show how it can
be reconciled with EPR results, but only if we accept that nature has
an ungraspable aspect, such as indeterminism or some other
form of uncontrollability, that prevents the violation of
strong locality from leading to faster-than-light signalling.

\begin{figure}[b]
\includegraphics[width=0.9\hsize]{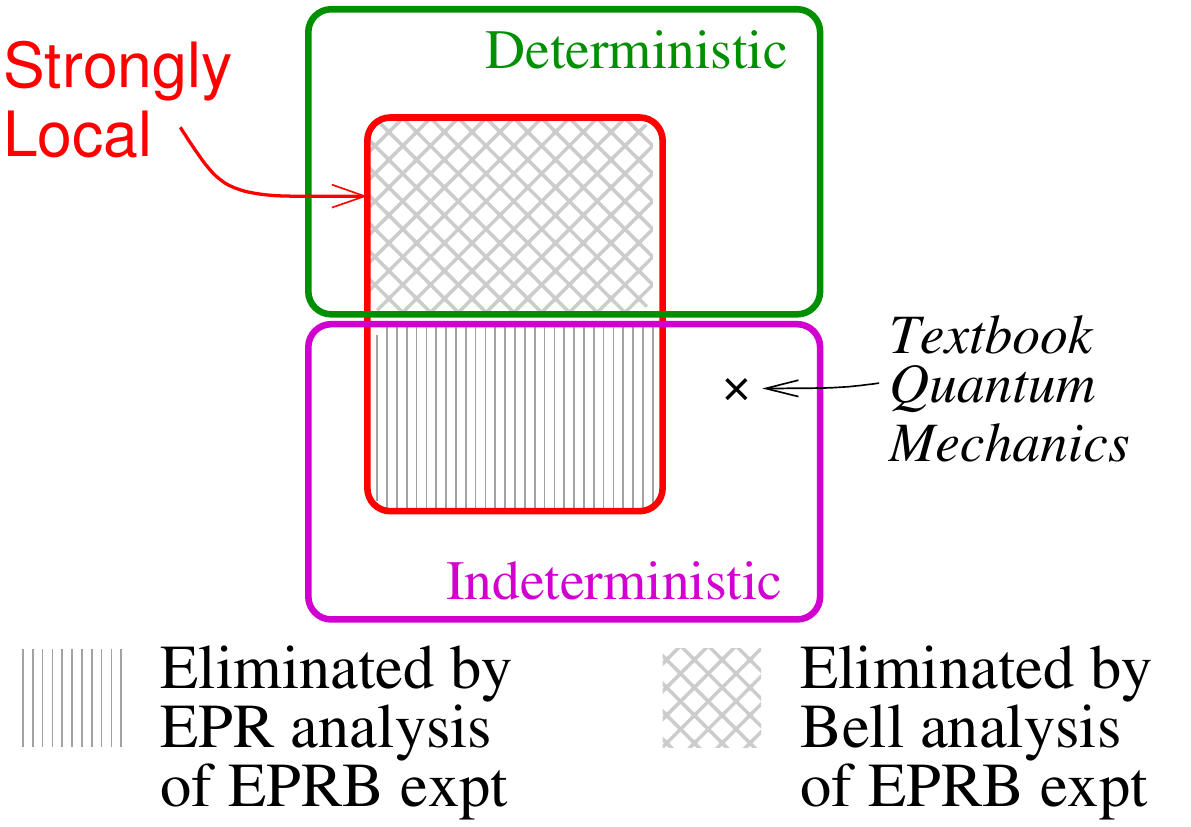}
\caption{Venn diagram of the space of theories and the constraints
from EPRB experiments.
The inner (red) rectangle encloses the set of strongly local theories.
The EPR analysis concludes that strongly local theories must be deterministic;
the Bell analysis concludes that strongly local theories cannot be deterministic.
In combination, these analyses rule out strongly local theories.
}\label{fig:locality-Venn-simple}
\end{figure}

\section{Overview}
\label{sec:overview}

% The ability of EPR experiments to test the principle of strong locality is
% expressed by Bell's inequality. Our simple non-mathematical explanation of
% this is in Sec.~\ref{sec:humans} and \ref{sec:EPR}.  We will show that Bell's
% inequality simply says that if you plan yes-or-no answers to three questions
% then on two randomly chosen questions your plan will lead you to give the same
% answer to both questions at least 1/3 of the time
% (Fig.~\ref{fig:answer_triangle}).

To make it clear how EPR experiments falsify the principle 
of strong locality, we now give an
overview of the logical context (Fig.~\ref{fig:locality-Venn-simple}).  
For pedagogical purposes it is natural to present the 
analysis of the experimental results
in two stages, which we will call the ``EPR analysis'', 
and the ``Bell analysis'' although historically they
were not presented in exactly this form \cite{Wiseman:2015wba,Norsen:2015};
indeed, both stages are combined in Bell's 1976 paper \cite{Bell:1976}.

We will concentrate on Bohm's variant of EPR,
the ``EPRB'' experiment. This involves pairs of particles, typically
a pair of photons in a spin singlet state. The question at hand is: 
what general types of theories can account for the observed behavior of
these particles? Can strongly local theories do the job?
Fig.~\ref{fig:locality-Venn-simple} shows the space of theories
of such particles. The inner (red) rectangle encloses the set of
strongly local theories, the ones
in which the probability of an event depends only on occurrences
in the event's past light cone.
% no causal influence can travel faster than light.
The upper (green) rectangle encloses the set of theories
that are ``deterministic'', meaning that
the behavior of the particles is is fully determined in advance
 without any randomness in the laws of physics.
When combined \cite{Bell:1976}, the EPR analysis and the Bell analysis show that no
strongly local theory, whether deterministic or indeterministic,
can explain the results of EPRB experiments.
We now give a brief outline of those analyses, to be expanded in later sections.

% As the figure shows, orthodox quantum mechanics (i.e. the form of quantum
% mechanics taught in conventional physics courses) is not a deterministic
% theory.  This is because in quantum mechanics the result of a measurement of
% the spin of a photon is not pre-determined by the state of the photon: the
% result is fundamentally random, with a probability distribution given by the
% wavefunction of the photon.  

%In Sec.~\ref{sec:uncertainty} we give a more
%detailed account of the meaning of strong locality and determinism.

\medskip\noindent {\bf The EPR analysis}\\
The EPR analysis \cite{Einstein:1935rr} 
starts with the experimental observation that both photons in the EPRB
setup show the same behavior when subjected to the same measurement,
no matter how far apart they are. The EPR analysis
then points out that if strong locality is 
true then this cannot be due to one photon influencing the other, so they must
have been pre-programmed to agree, which requires that
the photons have a deterministically-evolving
internal state that determines their behavior. In other words,
strong locality {\em requires} determinism
to explain the EPRB results.
The EPR analysis therefore rules out strongly local
theories that are indeterministic (vertical shading in
Fig.~\ref{fig:locality-Venn-simple}).  This sounds like a refutation of
quantum mechanics, which is famously an indeterministic theory.  However, 
``textbook'' quantum mechanics, as
taught in conventional physics courses, 
explicitly violates strong locality because
measurement induces instantaneous collapse of
the wavefunction over all of space, 
so EPR's analysis does not apply directly
to textbook quantum mechanics.  
Rather, it shows that any alternate theory that was strongly local
would have to be deterministic.  In such a theory the result of measuring a
photon's spin would not be random; it would be determined by the state
of non-quantum-mechanical ``hidden variables'' that predetermine the behavior
of the photon.

\medskip\noindent{\bf The Bell analysis and the Bell inequality} \\
The second stage of analysis of the
EPRB experimental data, which we call the ``Bell analysis'',
destroys the dream of
finding a strongly local and deterministic theory to replace quantum mechanics.
Bell pointed out that if nature is described by a strongly local and
deterministic theory then the behavior of the photon pairs 
has to obey a constraint called the ``Bell inequality''
\cite{Bell:1964kc,Bell:1976}.
In Secs.~\ref{sec:humans} and \ref{sec:EPR} we will 
give an elementary explanation of the Bell inequality
in terms of testing twins for faster-than-light telepathy.
We will show that it arises from
the fact that if someone has planned yes-or-no answers to three
questions then on two randomly chosen questions they will give the
same answer to both questions at least 1/3 of the time.

In real EPRB experiments (e.g.~\cite{Hensen:2015ccp})
the results violate Bell's inequality.
This shows that {\em no} deterministic and strongly local
theory can explain the behavior of the photons
(cross-hatched shading in Fig.~\ref{fig:locality-Venn-simple}).
Taken together, the EPR and Bell analyses of
the experimental data show that strong locality must be false.
If we accept the principle of common cause
(Sec.~\ref{sec:uncertainty})
this means that some causal influences travel faster than light.

The rest of this paper explores the EPR and Bell analyses in 
as much depth as is possible without mathematical formalism.
In Sec.~\ref{sec:uncertainty} we lay out in more detail the meaning
of the key postulates of strong locality and determinism.
In Sec.~\ref{sec:humans} we give an intuitive
non-mathematical explanation of the Bell inequality and the
resultant refutation of strong locality.
Sec.~\ref{sec:EPR} applies these concepts to the real experimental setup
involving photon spin measurements.

This paper focuses on strong locality because it is clear,
intuitively plausible, and can be cleanly defined
as a factorization condition (Eq.~\eqn{eqn:strong_locality}).
However, other analyses of the EPRB
experiment (e.g., \cite{Blaylock:2010,Maccone:2013}) do not use this definition,
and hence come to different-sounding conclusions about what
EPRB means for ``locality''.
In Sec.~\ref{sec:locality} we therefore explore
other principles that are related to locality, such as
``information cannot be transmitted faster than light''
or ``there is no preferred inertial reference frame'' (the Principle
of Relativity), and discuss how some form of locality might survive
even when strong locality is violated.
Sec.~\ref{sec:summary} gives a summary of our discussions.

\section{Locality and Determinism: definitions and assumptions}
\label{sec:uncertainty}

\begin{figure}
\includegraphics[width=0.9\hsize]{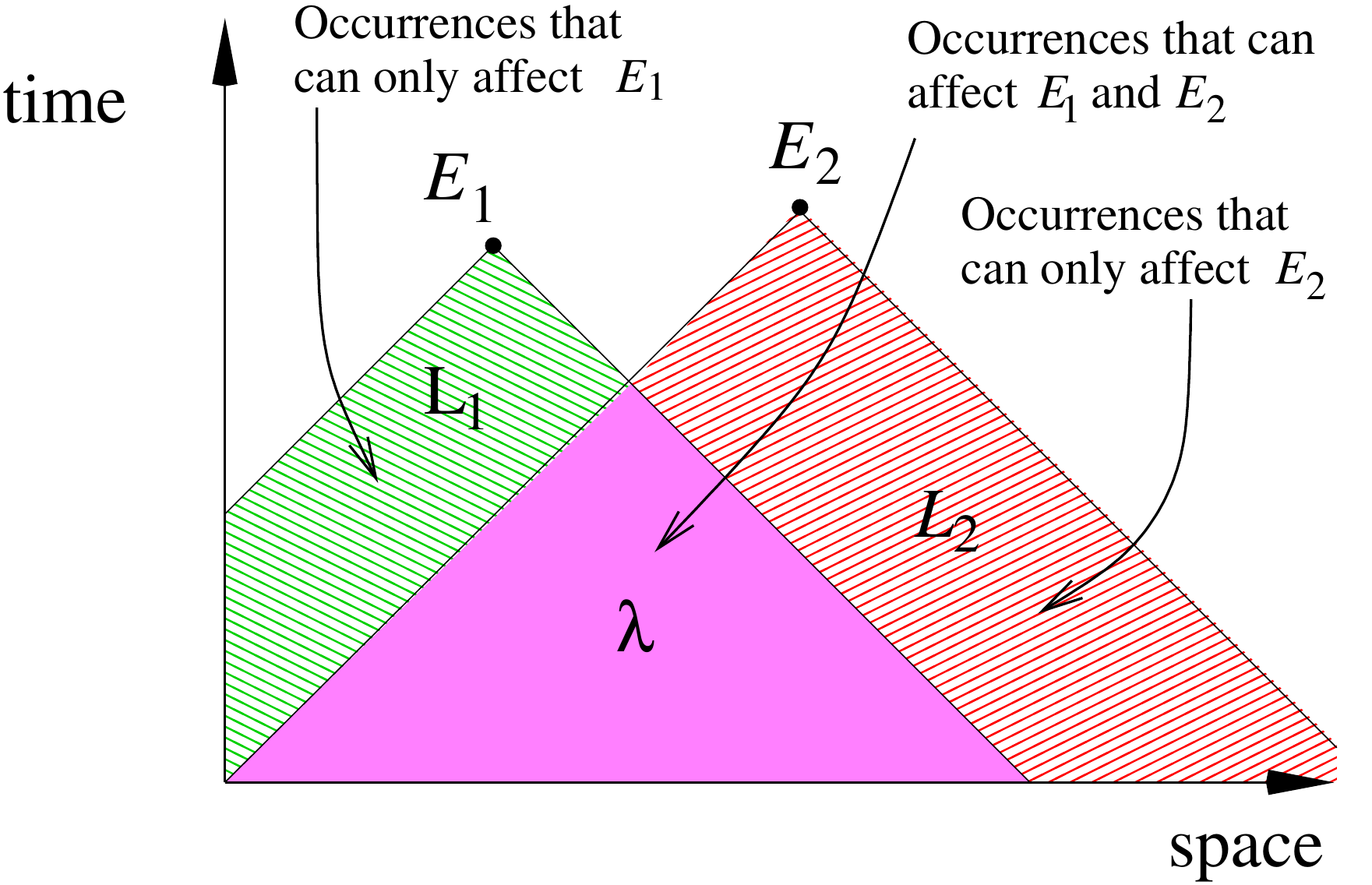}
\caption{Strong locality states that correlations between 
the results of two spacelike-separated measurements $E_1$ and $E_2$
can only arise from events in their shared past light cone.
}\label{fig:strong_locality}
\end{figure}

The principles that play a central role in
EPRB experiments are:

\ben
\item {\bf Determinism:} The result of any measurement on a
system is pre-determined by how the system was set up originally,
taking into account any subsequent influences on it. Any apparent
randomness just reflects our ignorance,
there is no essentially random component to the outcome
\cite{Jarrett:1989}.
In a deterministic theory, even for a measurement that 
was not actually performed
there is a fact of the matter about what result it would have yielded
(``counterfactual definiteness'').
\item {\bf Strong Locality:}
Once we take
into account everything in its past light cone,
the probability of an event
is not affected by additional information about things
that happened outside its past light cone
(Fig.~\ref{fig:strong_locality})  \cite{Bell:1976}.
This is sometimes called ``factorizability'' 
because it leads to a factorization of the 
probability function for 
space-like-separated events (Eq.~\eqn{eqn:strong_locality}).
%(Eqs.~\eqn{eqn:screening} and \eqn{eqn:strong_locality}).
As we will explain below, using a reasonable-seeming conception
of ``cause'' it is equivalent to 
saying that causal influences cannot travel faster than light, so
the causal influences that affect an event must be in its
past light cone.
\een

We now explain in more detail our background assumptions and
the meaning of determinism and strong locality.
Readers interested
in getting straight to the EPR and Bell analyses can skip
the rest of this section.

\medskip\noindent {\bf Background Assumptions}\\
In our discussion we will make the following background
assumptions. For a more fine-grained 
formulation see Ref.~\cite{Wiseman:2015wba}.
\ben 
\item ``Macro-realism'': each measurement has a unique outcome.
\item ``Random choices'':  each experimenter's choice of what to
measure is random, i.e.,
uncorrelated with the state of the particles being measured
and choices made by the other experimenter.
\een
These allow us to conclude from EPRB experiments that strong locality
is violated. To make a connection between strong locality and
causal influences, one needs
\ben
\setcounter{enumi}{2}
\item Reichenbach's principle of common cause \cite{Reichenbach:1956}: 
correlations can be explained in terms of causes.
if two phenomena show a correlation, either one causes the other or they
have a common cause. If $C$ is the common cause of $A$ and $B$ then
conditioning on $C$ factorizes the joint probability:
$p(A,B)=p(A|C)\, p(B|C)$.
\een
These assumptions
seem reasonable but not incontrovertible \cite{sep-qm-action-distance,sep-physics-Rpcc,tHooft:2001fb,Maudlin:2011,Wiseman:2015wba}. 
Proponents of many-worlds-type scenarios would deny macro-realism.
A superdeterminist or a believer in retrocausality
would not allow us to assume that the experimenters
choices can be treated as random.
Operationalists deny Reichenbach's principle.
We will comment further on these viewpoints in Sec.~\ref{sec:summary}.

\medskip\noindent {\bf Determinism}\\
Determinism states that the outcome of a measurement
is predetermined by the state of the system at earlier times,
taking into account any external influences on it.
In the context of EPRB experiments, as we will
see in Secs.~\ref{sec:humans} and \ref{sec:EPR}, determinism means that
the outcome of doing a measurement on a particle
can be reliably ``pre-programmed'' by
physical processes that set the initial states of two particle
before they are moved apart from each other.

% For example, imagine a particle, such as
% a cosmic ray, flying though space.  According to
% determinism, the particle has well-defined properties like its mass, speed,
% and amount of spin, and measurements expose these properties:
% the particle's speed has the value it has, whether or not you measure it.
% Moreover, those properties evolve in a deterministic way, 
% so with full knowledge of the particle and the environment through which
% it traveled, you would know how its properties will evolve over time.

Determinism is intimately bound up with
our understanding of {\em uncertainty}.
One can distinguish two ways in which we may be uncertain about the
outcome of a measurement:
\ben
\item Uncertainty arising from our ignorance. 
The outcome of the measurement could be predicted given
accurate knowledge of the initial state of the object and
the laws governing its evolution, 
but we don't have sufficiently accurate 
information about these things to make an exact prediction.
\item Fundamental uncertainty: the outcome of the
measurement has an essentially random component, either in the
evolution of the system or its effect on the measuring device.
In a sense the system gets to ``decide on its own'' how
to behave.
% until the measurement is actually performed.
\een

In ordinary life, and in science up until the advent of quantum mechanics, all
the uncertainty that we encounter is presumed to be of the first kind,
uncertainty arising from ignorance.  We can't predict the weather very
accurately, but the more we learn about the state of Earth's atmosphere and
oceans and the laws they obey, the better our predictions become.  Determinism
says that {\em all} uncertainty is of the first kind, the kind that arises
only from our ignorance.  Determinism is a sort of scientific optimism:
if we knew enough about the state of the universe
we could predict the outcome of any measurement.

Quantum mechanics introduced the idea that there might be uncertainty
of the second type, that nature might be fundamentally non-deterministic.
The EPR analysis shows that if strong locality is valid then
this sort of uncertainty is in conflict with the outcome of EPRB experiments.

\medskip\noindent {\bf Strong Locality}\\
%Strong locality states that
%Any correlation between two spacelike separated events
%can only arise from things in their shared past light
%cone that affected both events (Fig.~\ref{fig:strong_locality}).}
%once we take
%into account everything in its past light cone,
%the probability of an event
%is not affected by additional information about things
%that happened outside its past light cone.}
The application of strong locality
to the EPRB experiment is sketched in 
Fig.~\ref{fig:strong_locality}. Formally,
it says that any correlation between two spacelike separated events
$E_1$ and $E_2$ can only arise from 
each of them being correlated with events $\la$
in their shared past light cone.
Once we take into account those shared influences
the joint probability distribution of $E_1$ and $E_2$ factorizes
\cite{Jarrett:1989,Maudlin:2011,Wiseman:2015wba}:
\beq
p(E_1,E_2\,|\,L_1,L_2,\la) = p(E_1\,|\,L_1, \la)\times p(E_2\,|\,L_2, \la)
\label{eqn:strong_locality}
\eeq
where \\
\centerline{
\parbox{0.9\hsize}{
\newcounter{counta}
\begin{tightlist}{counta}{$\bullet$}
\item[$E_1$] is the outcome of the measurement on photon 1
\item[$E_2$] is the outcome of the measurement on photon 2
\item[$L_1$] is events in the past light cone of $E_1$ but not
$E_2$
\item[$L_2$] is events in the past light cone of $E_2$ but not
$E_1$
\item[$\la$] is everything in both light cones, or any other
state of affairs that can affect both $E_1$ and $E_2$
\end{tightlist}
}}

(Given strong locality, determinism is the statement that
$p(E_1\,|\,L_1, \la)$ is zero or one, and similarly for 
$p(E_2\,|\,L_2, \la)$.)

As we will see in Secs.~\ref{sec:humans} and \ref{sec:EPR}, for the EPRB
experiments to falsify strong locality each experimenter must decide ``at the
last minute'' what experiment to do on her photon. Thus the decision of what
measurement to perform on photon 1 occurs in $L_1$, so if strong locality is
true then that decision should not affect the measurement on photon 2, and
vice versa.  The assumption of random choices (Sec.~\ref{sec:uncertainty})
is crucial here; we assume that the choices made by the experimenters
are not influenced by the events $\la$ that determine the state of
the photons, hence the random choices assumption is often called
``lambda-independence'' \cite{sep-qm-action-distance}.

If we accept Reichenbach's principle of common cause 
then the violation of strong locality
means that there must be some faster-than-light causal influence that
allows the measurement of one photon to affect the measurement of the other
\cite{VanFrassen:1982,Norsen:2011,Wiseman:2015wba}.

\section{EPR and Bell with humans}
\label{sec:humans}

In order to make the EPR and Bell analyses of the EPRB data 
as comprehensible as possible
we now explain them using an analogy where instead of experimenting on photons
we are questioning people. For related approaches see, e.g.,
Sec 4.1.3 of Ref.~\cite{Preskill:1998}, or Ref.~\cite{Bricmont:2014}.

\subsection{Testing twins for superluminal telepathy}

Imagine that someone has told us that twins have special powers,
including the ability to communicate with each other using telepathic
influences that are ``superluminal'' (faster than light).
We decide to test this by collecting many pairs of twins, 
separating each pair,
and asking each twin one question to see if their answers agree.

To make things simple we will only have three possible questions, and they
will be Yes/No questions. We will tell the twins in advance what the
questions are.

The procedure is as follows.
\ben
\item A new pair of twins is brought in and told what the three possible questions are.
\item The twins travel far apart in space to separate questioning locations. 
\item At each location there is a questioner who selects one of the three 
questions at random, and poses that question to the twin in front of her.
\item {\bf Spacelike separation}: When the question is chosen and asked
at one location,
there is not enough time for any influence 
travelling at the speed of light to get
from there to the other location in time to affect either
what question is chosen there, or the answer given.
\een

\begin{figure*}[htb]
\parbox{\hsize}{
\includegraphics[width=0.9\hsize]{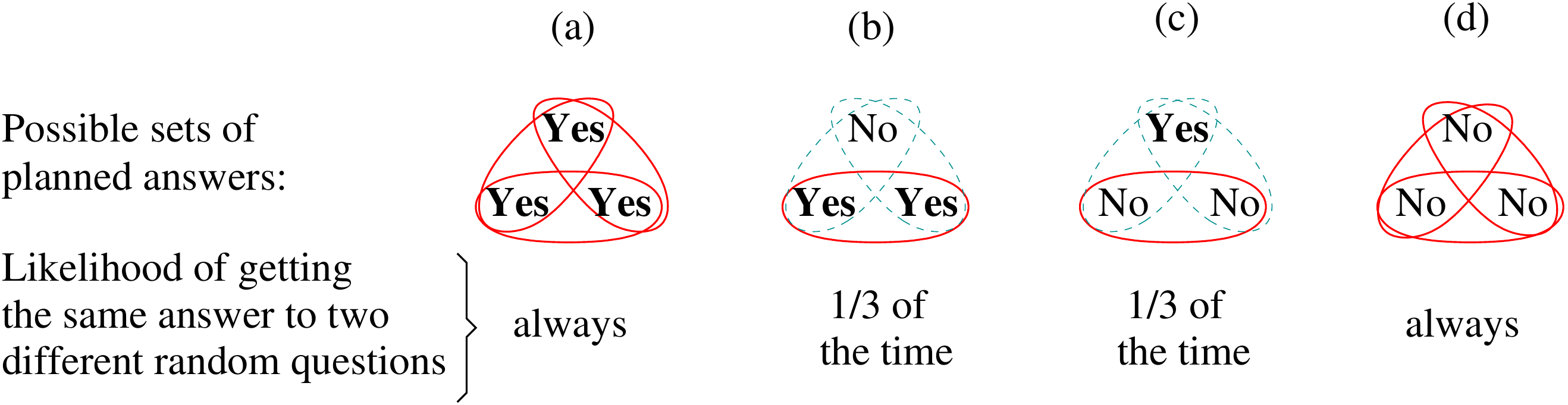}
}
\caption{
The essence of the Bell inequality 
(Eq.~\eqn{eqn:essence}). 
In formulating a plan for how to give
Yes/No answers to three questions, there
are four types of plan. 
No matter what plan one follows,
the answers to two different randomly chosen
questions will be the same {\em at least 1/3 of the time}.}
\label{fig:answer_triangle}
\end{figure*}

\subsection{EPR analysis of the data for the twins}

Now, suppose we perform this experiment and we find 
{\bf same-question agreement}: whenever
a pair of spacelike-separated twins
both happen to get asked the same question, their answers always agree.
How could they do this? 
There are two possible explanations,
\ben
\item Each pair of twins uses superluminal telepathic communication
to make sure both twins give the same answer.
\item Each pair of twins follows a plan. Before they were separated
they agreed in advance what their answers to the three questions would be.
\een
The same-question agreement that we observe does not
prove that twins can communicate telepathically faster than light.
If we believe that strong locality is a valid principle, then we can resort
to the other explanation, that each pair of twins is following a plan.
The crucial point is that this requires determinism. 
If there were any indeterministic 
evolution while the twins were
spacelike separated, strong locality requires that the random component
of one twin's evolution would have to be uncorrelated with the other twin's
evolution.  Such uncorrelated
indeterminism would cause their recollections of the plan to diverge,
and they would not always show same-question agreement.
This inference corresponds to the EPR analysis of
the EPRB experiment: strong locality (the
twins cannot exchange information faster than light), when combined with
same-question agreement, implies determinism 
(each pair of twins follows a predefined plan).

The idea that twins use a deterministically-evolving internal
``memory'' in order to follow a plan
 does not seem so remarkable, but for photons
this is a striking claim, because the quantum mechanical picture
of a photon does not allow for any internal state that determines
the outcome of measurements on a photon. 
The conclusion of
the EPR analysis (vertical shading in Fig.~\ref{fig:locality-Venn-simple})
is that if nature obeys strong locality then
only a deterministic theory can account for the agreement behavior seen
in EPRB experiments.

\subsection{Bell inequality for the twins}

In the thought experiment as described up to this point
we only looked at the recorded answers in cases where each twin
in a given pair
was asked the same question. There are also recorded data on
what happens when the two questioners happen to choose different questions.
Bell noticed that this data can be used as a cross-check on our 
strong-locality-saving idea
that the twins are following a pre-agreed plan that determines
that their answers will always agree.
The cross-check takes the form
of an inequality:
\beq
\parbox{0.9\hsize}{
\begin{flushleft}
\underline{Bell inequality for twins:}\\[1ex]
If a pair of twins is following
a plan then, when each twin  is
asked a {\em different} randomly chosen question,
their answers will be the same, 
on average, at least 1/3 of the time.
\end{flushleft}
}
\label{eqn:essence}
\eeq

Fig.~\ref{fig:answer_triangle} illustrates why \eqn{eqn:essence} is true. 
For each pair of twins, there are four
general types of pre-agreed plan they could adopt when they are
arranging how they will both give the same answer to
each of the three possible questions. 
\\
(a) a plan in which all three answers are Yes;\\
(b) a plan in which there are two Yes and one No;\\
(c) a plan in which there are two No and one Yes;\\
(d) a plan in which all three answers are No. \\
If, as strong locality and same-question agreement imply,
both twins in a given pair follow a shared predefined plan,
then when the random questioning leads
to each of them being asked a {\em different} question from the
set of three possible questions,
how often will
their answers happen to be the same (both Yes or both No)?
If the plan is of type (a) or (d), both answers will always be the same.
If the plan is of type (b) or (c),  both answers will be the same
1/3 of the time.
We conclude that
no matter what type of plan each pair of twins may follow, the mere fact that
they are following a plan implies that, when each of them is asked
a different randomly chosen question,
they will both give the same answer (which might be Yes or No)
at least 1/3 of the time (Eq.~\ref{eqn:essence}).
It is important to appreciate that one needs data from many
pairs of twins to see this effect, and that the inequality holds
even if each pair of twins freely chooses any plan they like.

This, then, is how the Bell analysis
applies to the data for the twins: strong locality 
(no way for the twins or questioners to influence each other when the
questioning is happening)
and determinism (each pair of twins follows a plan)
implies a Bell inequality \eqn{eqn:essence}.

\subsection{What if the twins violate the Bell inequality?}

In real experiments, when performing the analogous experiment on
photons, the Bell inequality is violated, showing that no
strongly local and deterministic theory can explain the data
(cross-hatched shading in Fig.~\ref{fig:locality-Venn-simple}).

Let us imagine the same thing happening in our analogy.
Suppose that when we analyze our results for a large sample of
twins, we find that in cases where each twin was
asked a different question, 
their answers are the same only 1/4 of the time;
3/4 of the time one twin gives a Yes and the other a No.
This result violates the Bell inequality \eqn{eqn:essence}, and
tells us that a good fraction of the population of twins
was {\em not} following any predefined plan when they answered the
questions. How do we interpret this result?

Our goal was to see if there was any evidence that the twins were
communicating with each other using telepathic influences
that travel faster than light.
The fact that the twins always agree when they are both asked the same
question, even when they
are being interrogated at spacelike separated locations,
could be explained away by assuming they were following a
prearranged plan. But if their pattern of answers 
to different questions violates
the Bell inequality then this shows that they can't be following
a prearranged plan. 
When one twin answers the question posed to him,
he needs to know what question his twin is being asked, because
if his twin is being asked a different question, at least 
some of the time one of them
will have to deviate from any pre-arranged plan,
changing his answer in such a way that it differs from the answer that 
his brother is giving, and thereby allowing their responses
to violate the Bell inequality.
Unless we are willing to discard one of the background assumptions
listed in Sec.~\ref{sec:uncertainty},
we are forced to accept that some sort of superluminal influence
connects the twins.

\begin{figure*}
\parbox{0.8\hsize}{
\includegraphics[width=0.8\hsize]{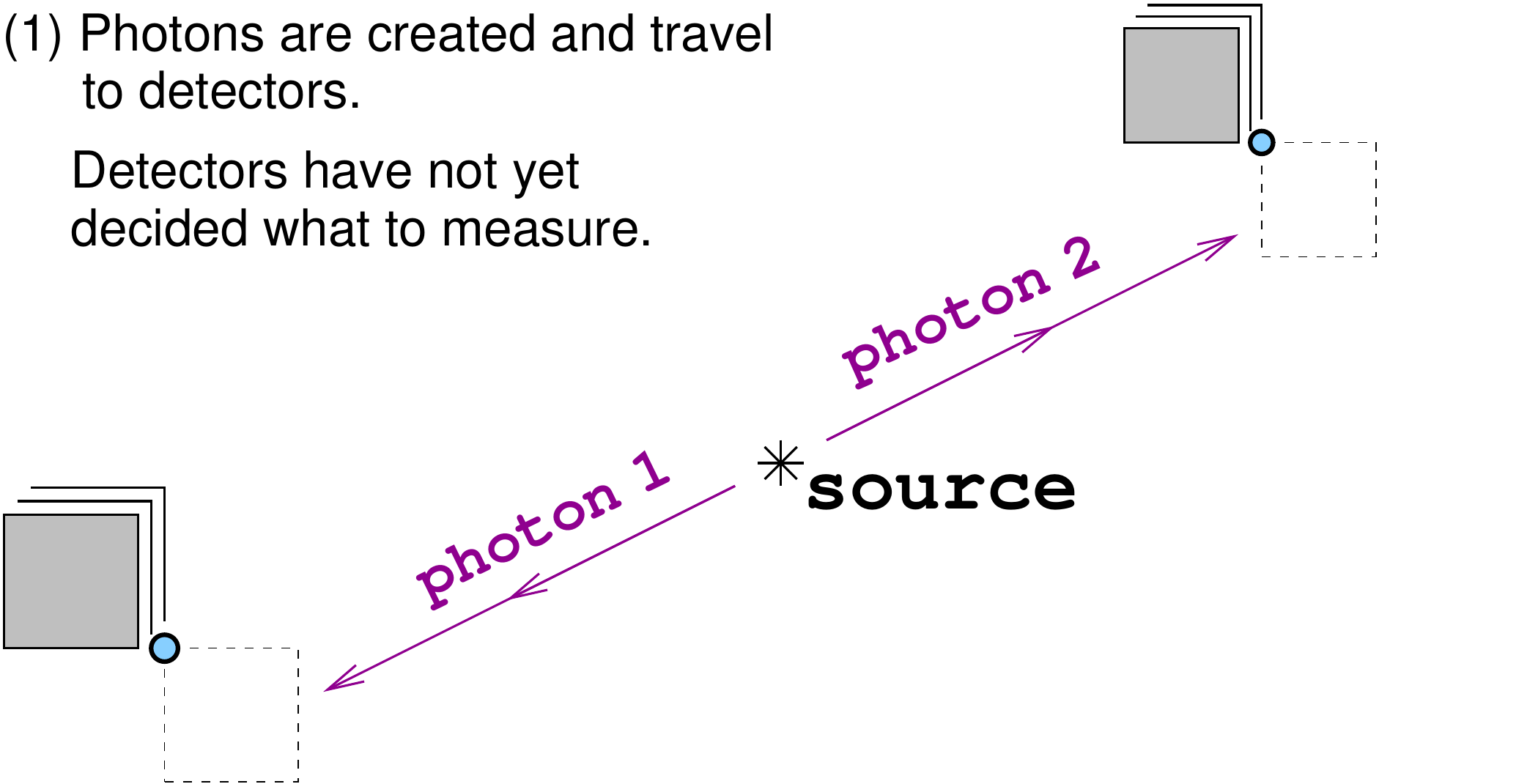}\\[8ex]
\includegraphics[width=0.8\hsize]{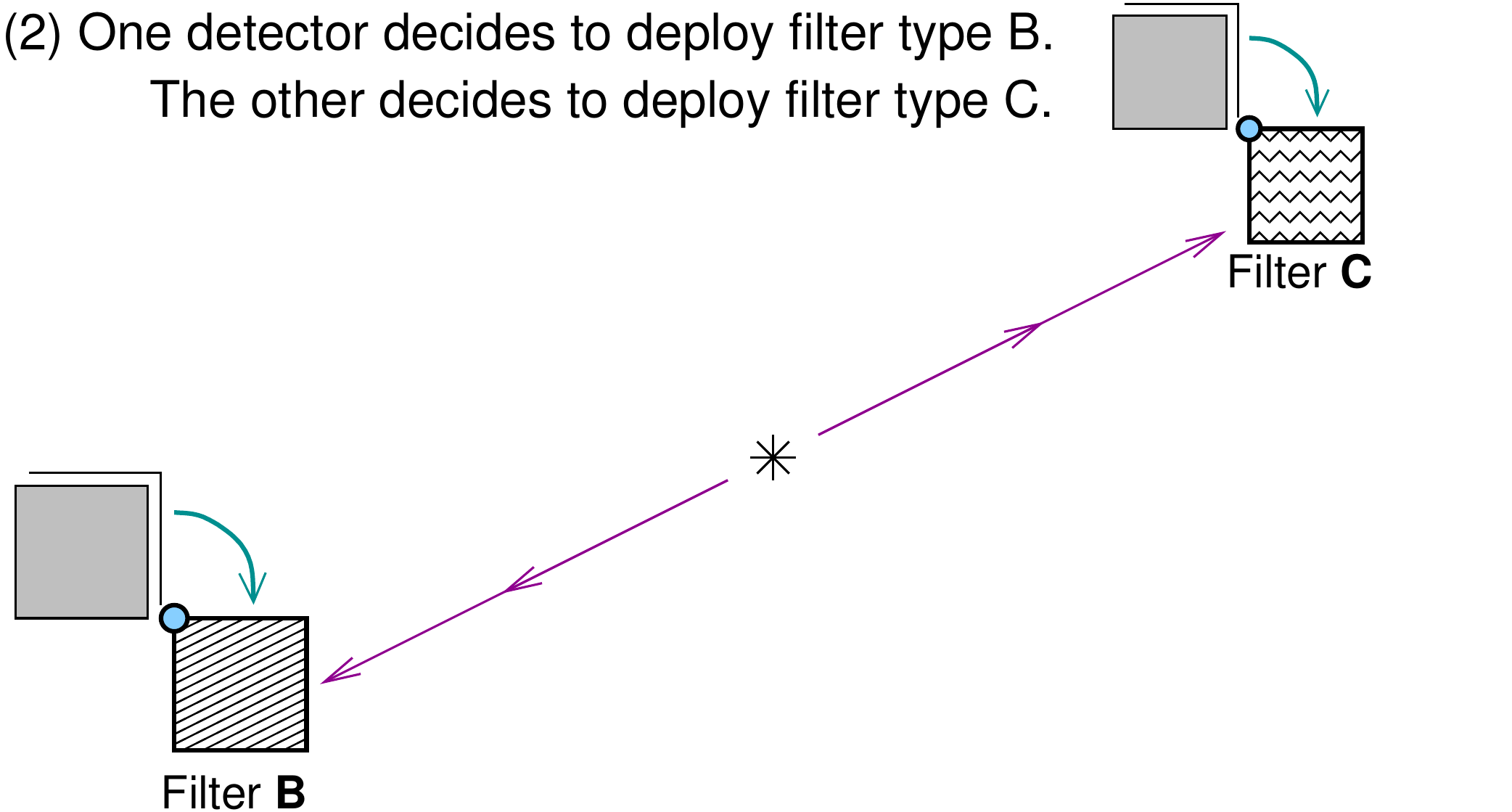}\\[8ex]
\includegraphics[width=0.8\hsize]{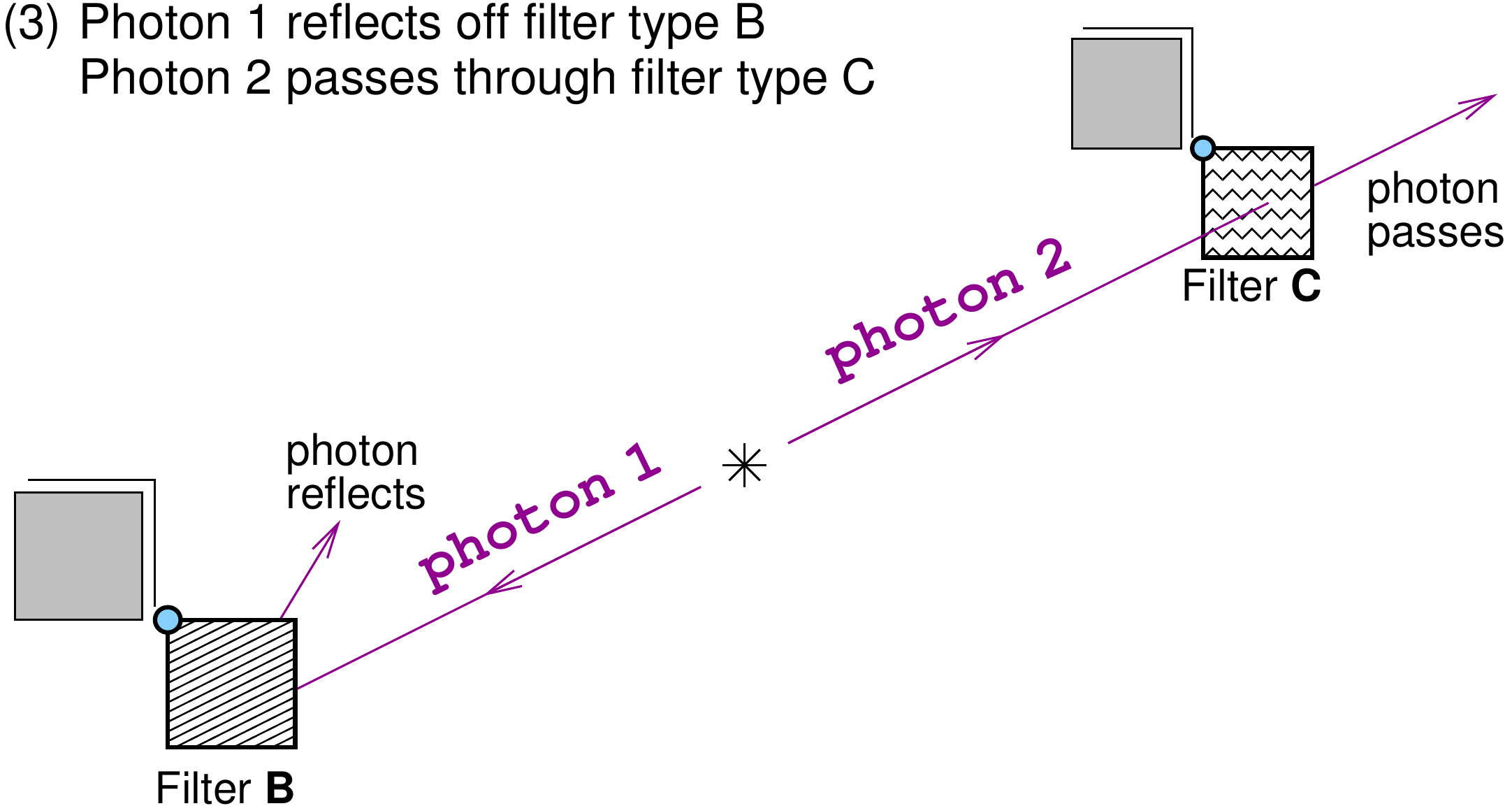}
}
\caption{One trial in the EPRB measurement of polarization for two photons.
The final result in this trial is 
% $(B\!:-,\, C\!:+)$, meaning 
that photon 1 encountered a filter of type B and reflected off it, while photon 2 encountered
a filter of type C and passed through it. According to the Bell inequality
(Eq.~\eqn{Bell-ineq}) this sort of result, where the two photons do
different things when encountering different filters, should happen
no more than 2/3 of the time.
}
\label{fig:EPR_polarizer}
\end{figure*}

\section{EPR and Bell with photons}
\label{sec:EPR}

The testing of twins for telepathic abilities, as described
in section \ref{sec:humans}, is an exact analogy to the EPRB
experiment, which is  a modification, suggested by Bohm \cite{Bohm:1951}, 
of the  original EPR experiment.
In the EPRB experiment (see Fig.~\ref{fig:EPR_polarizer})
there is a source that creates
pairs of photons, analogous to twins.
The photons travel out from the source in opposite 
directions. When they are far from each other, each photon
encounters a measuring machine that
can do three possible measurements. The machine contains
three types of filter, call them A,B, and C, and when the photon arrives
the machine flips one of the three types
of filter into the path of the photon. The photon has
two possible responses to the filter: it either goes
through the filter (``$+$'') or reflects off it (``$-$''). 
This is actually a measurement of the polarization of the
photon: each filter consists of polaroid with a different orientation
of its axis of polarization. 
If determinism is true then each photon has a
deterministically evolving inherent polarization state that
determines how it will interact with each filter.

%The EPRB experiment is therefore exactly analogous to the situation 
%described in section
%\ref{sec:humans}, where each twin gave a yes-or-no answer and there were three
%possible questions he might be asked; in that case determinism corresponded to
%the twins having a plan that determined how they would answer each question.

If both machines deploy the same filter
then we see ``same-axis agreement'': either both photons  pass through
or they both reflect off.

%$(B\!:+, B\!:+)$  $(B\!:-, B\!:-)$. 
As with the twins, we can immediately see two ways to explain this consistent
agreement.
\ben
\item {\em Influence}: 
when one photon reaches its machine and 
the machine decides what filter to flip up in front of it and
the photon responds to that filter, 
some information is superluminally transmitted to
the other photon so that
if the other photon gets the same filter, it will behave the same way.
\item {\em Determinism}: when the photons are created, 
each is formed in a state (its ``polarization state'')
that determines how it will respond to any
possible filter it might encounter. The source puts
both photons into the same state, and those states
evolve deterministically,
ensuring that the photons always behave the same way when they
encounter the same type of filter.
\een
The EPR analysis (vertical shading in Fig.~\ref{fig:locality-Venn-simple})
concludes that in any strongly local theory, since there are no
faster-than-light correlation-creating influences,
agreement in same filter (same axis) measurements must arise
from the photons having a deterministically evolving internal state 
that pre-determines
their response to the filters that they encounter.
If, as EPR did, one takes strong locality to be valid, then 
the observed same-axis agreement
shows that the photons are in a state that determines their
behavior, which is in contradiction with the quantum mechanical
picture where their state does not determine the outcome
of measurements performed on them.

However, just as for the twins, there is a Bell analysis 
(cross-hatched shading in Fig.~\ref{fig:locality-Venn-simple})
which shows that EPR's picture, of physical objects
having deterministically evolving states and strongly local interactions, can be experimentally
tested. For this we look at the data for the cases when the two measuring
machines deploy {\em different} filters in front of the two photons.
Following the logic used in Sec.~\ref{sec:humans}, 
we conclude that if both photons are in the same polarization state,
and there is no correlation-creating 
influence between their spacelike-separated measurements, then,
on the occasions that
the detectors deploy {\em different} filters, then photons 1
and 2 should show the same behavior (both bouncing off or both passing
through) at least 1/3 of the time:
\beq
\parbox{0.9\hsize}{
 \mbox{{\bf Bell inequality:}}\\[1ex]
 \fbox{
  \mbox{prob}$\biggl($\bt{l}
   \mbox{when photon 1 and photon 2}\\[-0.5ex]
   \mbox{encounter different filters,}\\[-0.5ex]
   \mbox{they show the same behavior} 
  \et 
  $\biggr)\geqslant 1/3$
 }
}
\label{Bell-ineq}
\eeq
In Appendix~\ref{sec:Bell} we show
how Eq.~\eqn{Bell-ineq} is a form
of Bell's original inequality.

When polarizations of pairs of spin-singlet photons are measured in
real-world experiments, it is found that they do show 
agreement in same-axis measurements, but when we perform different-axis
measurements the two photons only show the same behavior 1/4 of the time;
3/4 of the time they show different behavior: one bounces off its filter
and the other passes through.
This violates the Bell inequality.
Such violation has now been seen in many experiments,
e.g.~\cite{Gisin:2008,Zeilinger:2013,Hensen:2015ccp}.

We conclude that strong locality is violated by spin-singlet photon pairs.
Either you need a strong-locality-violating influence to make the
same-axis agreement happen, or, if you try to save
strong locality by assuming that each photon is in a
state (the same state for both of them) 
that pre-determines the outcome of measurements on it,
then you need a strong-locality-violating influence to
obtain the observed
violation of Bell's inequality for different-axis measurements.
Either way, a
violation of strong locality is required to account for all the relevant
experimental observations.

\section{Consequences for Locality}
\label{sec:locality}

The EPRB experiment, in combination with some assumptions that
we have outlined in Sec.~\ref{sec:uncertainty}, tells us  that
nature does not obey the principle of strong locality.
If we accept Reichenbach's principle of common cause
we would say that there are causal influences that travel faster then light.
But this cannot be the end of the story:\\
$\bullet$ What about Einstein's theory of relativity? Are EPRB results
compatible with the Principle of Relativity?\\
$\bullet$ If so, is there
some ``medium-strength'' locality principle, implied by Relativity
but weaker then Strong Locality, that
is compatible with EPRB experiments? \\
$\bullet$ What about determinism? Do the EPR and Bell analyses leave
open the possibility of deterministic theories?

We will now explain why it is believed that EPRB experiments do not violate
the Principle of Relativity,
and suggest that ``signal locality'' is a useful
medium-strength locality principle, since it distills the requirements
of relativity and chronology protection 
(the absence of causal paradoxes \cite{Hawking:1991nk}).
We will acknowledge that signal locality contains concepts such as ``control''
that are not usually present in physical principles,
and argue that, although signal locality
is compatible with determinism, nature must
have some inherent elusiveness,
perhaps indeterminism or perhaps some form of
uncontrollability, in order for the
EPRB results to be consistent with signal locality and hence
with Relativity and chronology protection.

\subsection{EPR and the Principle of Relativity}
\label{sec:relativity}
%We finally come to the question of how EPRB experiments 
%cohere with the Principle of Relativity.
To quote Bell himself,
{\em ``one of my missions in life is to get people to see that if they want to talk about the problems of quantum mechanics---the real problems of quantum mechanics---they must be talking about Lorentz invariance''}
\cite{sep-qm-bohm}.
In this quote, ``Lorentz invariance'' is just the Principle of Relativity, 
which states
that the laws of physics are the same in all inertial reference frames,
so the laws of physics
are invariant under the Lorentz transformations that relate
different reference frames to each other.

%Bell is concerned that the Principle of Relativity
%might turn out to imply strong locality, in which case
%EPR experiments, by refuting strong locality,
%would actually refute relativity itself \cite{Norsen:2008}.
%the ``allowed region'' in  Fig.~\ref{fig:locality-Venn} (of theories
%that obey signal locality and the EPR/Bell constraints) would be
%incompatiblesubset that obeys the Principle of Relativity.

So, is the faster-than-light
connection between distant photons that we see in EPRB
experiments
compatible with the Principle of Relativity?
There is evidence that they are compatible, but not in the
straightforward way that one might assume.

Naively one might say that of course the EPRB results are consistent with 
the Principle of Relativity, because they agree with 
the predictions of quantum mechanics,
and there is a relativistic, Lorentz-invariant, formulation of 
quantum mechanics, namely quantum field theory.
%This might seem like an existence proof that the EPRB results are 
%compatible with Relativity.
It is true that most presentations of quantum field theory 
seem Lorentz invariant because they
focus on expectation values and do not
discuss the measurement postulate
(instantaneous wavefunction collapse induced by the measurement process).
But textbook quantum mechanics, including quantum field theory, needs
the measurement postulate to explain how unique experimental results
arise from measurements (the ``macro-realism'' assumption
of Sec.~\ref{sec:uncertainty}, discussed further in Sec.~\ref{sec:summary}).
There is no Lorentz-invariant
version of measurement-induced wavefunction collapse
that is compatible with the EPRB results \cite{Maudlin:2011}.
However, this does not rule out the possibility that there may be
other Lorentz-invariant theories that can explain the EPRB results.
In fact, in 2006 an example was proposed: a version of quantum mechanics
where the wavefunction occasionally
collapses spontaneously in a Lorentz-invariant way
\cite{Tumulka:2006}.
Whether or not this theory is a correct description of nature, 
it seems to
provide an existence proof that EPRB results are compatible
with the Principle of Relativity.

\subsection{Different forms of locality}
\label{sec:forms}

If the Principle of Relativity is compatible with
EPRB experiments but strong locality is not, then strong locality
does not follow from the Principle of Relativity. 
However, there is another locality principle, {\em signal locality},
that does plausibly follow from relativity combined with chronology
protection (no causal paradoxes).
To set the context for our discussion of
signal locality, here is a quick survey of
various requirements that can be thought of as expressing
ideas of locality, 
along with a summary of how compatible they
are with EPRB experiment results: 
\begin{tightlist}{counta}{\arabic{counta})}
\item {\em Strong locality} (or {\em local causality}): 
after taking into account everything in its past light cone,
the probability of an event
is not affected by additional information about things
that happened outside its past light cone.\\
As we have seen, this is disproven by EPRB experiments.
\item {\em Information} must be transmitted no faster than light.\\
This is also disproven by EPRB experiments, since the
result of the measurement on one photon contains information 
about the measurement performed on the other that did not come
from the backward light cone.
\item {\em Signal locality}: signals can travel no faster than light.\\
This is compatible with EPRB experiments, but at a 
price, as we will describe below.
\item {\em Energy or other conserved quantities} must travel 
no faster than light.\\
This is compatible with EPRB experiments, since 
there is no evidence that any physical substance
travels from one photon's measurement site to the other's.
\item {\em The Principle of Relativity}: The laws of physics 
are the same for any observer who is not accelerating (any ``inertial
frame of reference'').\\
As discussed above, this is compatible with EPRB experiments.
\end{tightlist}

\smallskip

\begin{figure}
\includegraphics[width=\hsize]{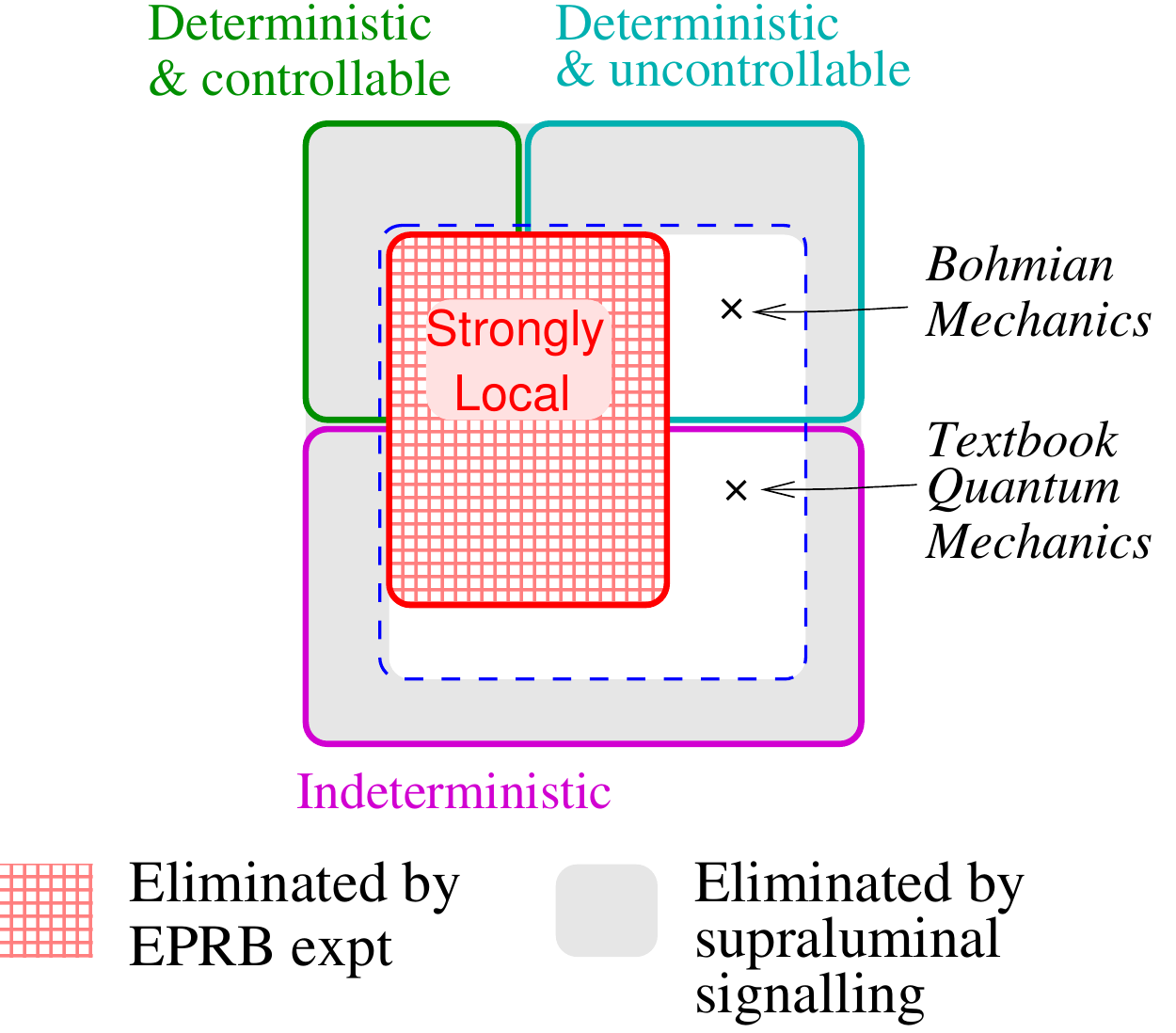}
\caption{Augmented version
of Fig.~\ref{fig:locality-Venn-simple}, showing the
set of theories obeying signal locality (enclosed by the
dashed (blue) line), and dividing deterministic theories into
those where the hidden variables can be controlled
and those where they cannot.
The grey shaded region is the set of theories where
superluminal signalling is possible. Theories in the white area
are consistent with EPRB experiments and have
the right kind of indeterminism/uncontrollability to be
consistent with signal locality.
For details see Sec.~\ref{sec:signal-locality}.
}
\label{fig:locality-Venn-signal}
\end{figure}

\subsection{EPR and Signal locality}
\label{sec:signal-locality}

In discussing signals, the essential point is that signalling is more than
the transfer of information. Sending a signal means having a 
{\em controllable} means of transferring information.
% Also, can non-Lorentz-invt theories ( like QM with instantaneous
% collapse) show signal locality?
Control, however, is based on high-level 
concepts such as agency and free will, and such concepts
are not usually invoked in fundamental physical principles.
Bell complained that signal locality ``rests on
concepts which are desperately vague, or vaguely applicable. The
assertion that `we cannot signal faster than light'
immediately provokes the question: Who do we think {\em we} are?''
\cite{Bell:1990}.

In view of this concern, we will proceed by treating
signal locality as a property (of theories) that we hope will be
implemented by some more fundamental feature of the theory.
Theories with the property of signal locality are attractive
because they can obey the Principle of Relativity
(no preferred inertial frame of reference) while  
maintaining chronology protection, i.e.~avoidance of causal paradoxes.
There is danger of a causal paradox if someone can
send a signal to themselves in the past,
since the person could, after receiving the signal, decide 
(assuming free will) not to send it. (For a discussion
not assuming free will see Ref.~\cite{sep-time-travel-phys}.)
To make sure that this cannot happen, we have to ensure that
the sender of controllable information (signals) is always in the
past of the receiver. In a theory that obeys the Principle of Relativity,
this means that
signals must go slower than light, since only then will
all reference frames agree on the time ordering of the sender and receiver.

We can now see that EPRB experiments, which show that information
can be transferred faster than light, also require some accompanying
element of uncontrollability in nature in order to make sure that 
this does not translate to a violation of
signal locality. We can imagine two ways to preserve signal locality
in the face of the results of EPRB-type experiments.\\[1ex]
(a) {\em Nature is indeterministic.}\\
The measurement outcomes are uncontrollable because nature is
fundamentally indeterministic. 
In an indeterministic theory there is a random component to the
evolution and/or measurement of the photons, so
we have to rely on some sort of faster-than light influence to 
produce the consistent agreement in the results of 
space-like separated same-axis measurements
and also to produce the different-axis correlations that
violate the Bell inequality.
But this does not allow superluminal signalling because the measurement results
are not determined by anything: they are inherently random
and therefore uncontrollable. In this case signal locality
follows from the requirement of
``parameter independence'' (see Appendix~\ref{sec:remote}).
Textbook quantum mechanics is an example of an indeterministic 
parameter-independent theory. The instantaneous collapse
of the wavefunction is the faster-than-light influence that
ensures same-axis agreement. \\[1ex]
%The information that is transferred from one measurement location to the other
%is information about the result
%of the measurement which arises randomly from the collapse of the
%wavefunction. Since it is uncontrollable there is no problem
%with this information being transferred faster than light. It cannot be used to
%send a signal.\\[1ex]
%
(b) {\em Nature is deterministic but uncontrollable.}\\
The measurement outcomes are uncontrollable because, even though
they are determined by the states of the objects in question
(``hidden variables''), those states are themselves 
sufficiently uncontrollable,
because of some physical law, that they cannot be used to send signals.
In such a scenario the violation of Bell's inequality in the
different-axis measurements arises from
a faster-than-light influence that allows one measurement to affect
the hidden variables of the object being measured at the other,
space-like separated, location, but the experimenter at one location
can never control the hidden variable states well enough to be able to
control the measurement results at the other end, and thereby send a
message.
Bohmian Mechanics is an example of a theory that follows this pattern
\cite{sep-qm-action-distance,Maudlin:2011}.
%Experimental inaccessibility of the hidden variables may also be an important
%factor in ensuring that they cannot be used for superluminal signalling,
%but to the degree that they have faster-than-light effects
%on the results of remote measurements, they must be uncontrollable.

In Appendix~\ref{sec:remote} we describe Jarrett's more formal way of
exposing this dichotomy, by analyzing strong locality into two
weaker conditions \cite{Jarrett:1989}. 
Violation of one of them (``Remote Outcome Independence'')
corresponds to possibility (a) above; violation of the other
(``Parameter Independence'' or ``Remote Detector Independence'') 
corresponds to possibility (b).

In Fig.~\ref{fig:locality-Venn-signal} we show an augmented version of
the simple theory-space diagram (Fig.~\ref{fig:locality-Venn-simple}),
including the set of theories that obey signal locality (enclosed by the
dashed (blue) line). If signal locality is valid then
physics is restricted to the unshaded (white)  area 
of allowed theories.
These all violate strong locality, as required by 
the EPRB experimental results, but in a way that avoids
superluminal signalling as described above.

\section{Summary}
\label{sec:summary}

The EPRB experiment uses spin-singlet photon pairs to test the
degree to which the laws of nature obey some sort of principle of locality.
%
% \iffalse
% By separately measuring the spins of the two photons when they are spacelike
% separated, so only faster-than-light influences could allow the measurement
% of one photon to affect the measurement of the other, we find\\
% $\bm{-}$ {\em same-axis agreement}: when we measure the spins of 
% both photons along the same axis, the two results are always the same.\\
% $\bm{-}$ {\em Bell-violating levels of different-axis agreement}:
% when we measure the spins of both photons along different axes, they
% agree less often than they would if the result were determined by
% some internal state of each photon.\\
% \fi
%
If we accept some background assumptions (Sec.~\ref{sec:uncertainty})
then the EPRB experimental results bring us to the following conclusions.
\bi
\item The observed behavior violates the principle of
strong locality (local causality) which states that
%once we take into account everything in its past light cone,
%the probability of an event
%is not affected by additional information about things
%that happened outside its past light cone
no correlation-creating
influence can travel faster than light.
In a nutshell, this is because in the EPR experiment
either you need a faster-than-light 
influence to make the
same-axis agreement happen, or, if you try to save
strong locality by assuming that the agreement arises from the photons
being in states that determine in advance that their spins will have
specific values, then you need a faster-than-light influence to
obtain the violation of Bell's inequality for different-axis measurements.
\item The EPRB results are % believed to be 
compatible with the Principle of Relativity
(equivalence of all inertial reference frames, also called Lorentz invariance).
\item In order to avoid casual paradoxes we expect nature to 
display the property of {\em signal locality} 
(signals cannot travel faster than light). 
This means there must be sufficient indeterminism or
uncontrollability in nature to prevent the EPRB correlations from being used for signalling.
\ei

If we prefer to explain the EPRB results by adopting an
indeterministic theory (such as textbook quantum mechanics) then,
while accepting that strong locality is violated, we can ensure that
the theory is signal-local by imposing a
weaker locality principle like ``parameter independence'', 
although, as described in Appendix~\ref{sec:remote},
the concept of parameter independence also involves non-fundamental
concepts of the type that made Bell object to signal locality as a
fundamental principle.
Treatments of EPRB that favor this approach
(e.g.~\cite{Blaylock:2010,Maccone:2013})
tend to de-emphasize the violation of strong locality and 
frame EPRB as forcing us to choose between
determinism and a weaker form of locality.

If we wish to preserve determinism then we must come up with a
theory (such as Bohmian mechanics) in which there are superluminal
influences between the deterministically evolving hidden variables,
but we face the challenge of constructing the theory so that 
it preserves signal locality by enforcing
essential limits on our ability to control those variables,
so they cannot be used for signalling.

In Sec.~\ref{sec:uncertainty} we listed the background assumptions
used in our analysis. We now briefly discuss the possibility
of dropping those assumptions. 
For a fuller discussion see Ref.~\cite{Wiseman:2015wba}.

Dropping the assumption of macro-realism
(experiments have unique outcomes) 
%(Sec.~\ref{sec:relativity}),
renders our entire analysis moot. However, an anti-macro-realist
must then explain how it is that experiments always appear to have
unique outcomes. Anti-macro-realist versions of quantum mechanics
such as the  many-worlds \cite{Everett:2012}
or many-minds  \cite{Albert:2009}
interpretations lead to questions of how probabilistic
predictions emerge and the role of decoherence \cite{sep-qm-action-distance}.

It is possible to deny that the choices made by the experimenters
can be treated as random. For example, 
a superdeterminist (e.g.~\cite{tHooft:2001fb})
would suggest that some mechanism ensures that those choices 
are always predetermined just so as to violate the Bell
inequality. This seems difficult, given that there 
are many ways to design a pseudo-random number generator for each experimenter
to use, including ones that are sensitive to events outside their
shared past lightcone $\la$ (see Fig.~\ref{fig:strong_locality}
and Ref.~\cite{sep-qm-action-distance}).
Alternatively, a believer in retrocausality would suggest that the 
experimenters'
choices could exert influence backwards in time on the state
in which the particles were prepared. This calls for an
explanation of why such retrocausality does not lead to violations of chronology
protection \cite{Pegg:2008}.

Reichenbach's principle of common cause is not an essential
assumption but it plays a fundamental role in science.
An ``operationalist'' or ``instrumentalist'' would
reject it  \cite{VanFrassen:1982,Fine:1989}, admitting that
EPRB experiments violate
strong locality, but maintaining that not all correlations can be explained
in terms of causes that factorize correlations (see Sec.~\ref{sec:uncertainty}), so
this is not a sign of superluminal causal 
influences. However, since most of science consists of the search for
the causes of correlations, the operationalist then has to explain
which correlations call for such causal explanations and which do not.
A Quantum Information theorist would also reject Reichenbach's principle,
claiming that quantum entanglement
can cause correlations without causing the individual events that 
exhibit the correlation \cite{Maccone:2015}.

In conclusion, the EPRB experiment exposes some of the complexity
of the concept of locality. Strong locality, which seems
simple and intuitively attractive, is violated by the
results while the Principle of Relativity is not. Signal locality
can be preserved, but it is formulated in terms of concepts like
``control'' which seem out of place in a theory of physics.

Quantum mechanics uses indeterminism to avoid superluminal signalling
and impressively accounts for the
unusual characteristics of the EPRB correlations
(they are not attenuated with distance, and they
only connect specific particles that were created in entangled
states) but in textbook quantum mechanics the
measurement-induced collapse of the wavefunction is 
instantaneous over all space and therefore not Lorentz invariant,
so one natural goal is to find and empirically validate a
Lorentz invariant form of wavefunction collapse such as that proposed
by Tumulka \cite{Tumulka:2006}.

The EPRB results also leave open the possibility of theories that,
unlike quantum mechanics, are deterministic, with 
signal locality ensured by limits
on the controllability of the hidden variables. Bohmian mechanics
is a well known proposal, and there is an ongoing
search for a Lorentz-invariant version of it \cite{Durr:2013asa}.

\bc{\bf Acknowledgments}\ec
I thank Joseph Berkowitz, Jeffrey Bub, Carl Craver, Tim Maudlin, Travis Norsen,
Kasey Wagoner, Howard Wiseman, and three anonymous referees
for helpful comments.

% Override the revtex href command in order that the JHEP bib style
% will work properly:
\renewcommand{\href}[2]{#2}
\bibliographystyle{JHEP_MGA}
\bibliography{EPR_and_Bell} 

\providecommand{\href}[2]{#2}\begingroup\raggedright\begin{thebibliography}{10}

\bibitem{Hensen:2015ccp}
B.~Hensen {\em et.~al.}, {\it {Experimental loophole-free violation of a Bell
  inequality using entangled electron spins separated by 1.3 km}},  Nature {\bf
  526} (2015) 682--686, [\href{http://xxx.lanl.gov/abs/1508.0594}{{\tt
  arXiv:1508.0594}}].

\bibitem{Bell:1976}
J.~S. Bell, {\it {The theory of local beables}},  Epistemological Lett. {\bf 9}
  (1976) 11--24.

\bibitem{Maudlin:2011}
T.~Maudlin, {\em Quantum Non-Locality and Relativity: Metaphysical Intimations
  of Modern Physics}.
\newblock Wiley, 2011.

\bibitem{Norsen:2015}
T.~{Norsen}, {\it {Are there really two different Bell's theorems?}},  ArXiv
  e-prints (Mar., 2015) [\href{http://xxx.lanl.gov/abs/1503.0501}{{\tt
  arXiv:1503.0501}}].

\bibitem{Wiseman:2015wba}
H.~M. Wiseman and E.~G. Cavalcanti, {\it Causarum investigatio and the two
  bell's theorems of john bell},  2015.
\newblock \href{http://xxx.lanl.gov/abs/1503.0641}{{\tt arXiv:1503.0641}}.

\bibitem{Einstein:1935rr}
A.~Einstein, B.~Podolsky, and N.~Rosen, {\it {Can quantum mechanical
  description of physical reality be considered complete?}},  Phys.Rev. {\bf
  47} (1935) 777--780.

\bibitem{Bell:1964kc}
J.~Bell, {\it {On the Einstein-Podolsky-Rosen paradox}},  Physics {\bf 1}
  (1964) 195.

\bibitem{Blaylock:2010}
G.~Blaylock, {\it The epr paradox, bell's inequality, and the question of
  locality},  American Journal of Physics {\bf 78} (2010), no.~1 111--120.

\bibitem{Maccone:2013}
L.~Maccone, {\it A simple proof of bell's inequality},  American Journal of
  Physics {\bf 81} (2013), no.~11 854--859.

\bibitem{Jarrett:1989}
J.~Jarrett, {\it Bell's theorem: a guide to the implications},  in {\em
  Philosophical Consequences of Quantum Theory: Reflections on Bell's theorem}
  (J.~T. Cushing and E.~McMullin, eds.), pp.~60--79.
\newblock University of Notre Dame Press, 1989.

\bibitem{Reichenbach:1956}
H.~Reichenbach, {\em The Direction of Time}.
\newblock Berkeley: University of Los Angeles Press, 1956.

\bibitem{sep-qm-action-distance}
J.~Berkovitz, {\it Action at a distance in quantum mechanics},  in {\em The
  Stanford Encyclopedia of Philosophy} (E.~N. Zalta, ed.).
\newblock fall 2013~ed., 2013.

\bibitem{sep-physics-Rpcc}
F.~Arntzenius, {\it Reichenbach's common cause principle},  in {\em The
  Stanford Encyclopedia of Philosophy} (E.~N. Zalta, ed.).
\newblock fall 2010~ed., 2010.

\bibitem{tHooft:2001fb}
G.~'t~Hooft, {\it How does god play dice? (pre)determinism at the planck
  scale},  \href{http://xxx.lanl.gov/abs/hep-th/0104219}{{\tt hep-th/0104219}}.

\bibitem{VanFrassen:1982}
B.~C. Van~Fraassen, {\it The charybdis of realism: Epistemological implications
  of bell's inequality},  Synthese {\bf 52} (1982), no.~1 25--38.

\bibitem{Norsen:2011}
T.~{Norsen}, {\it {John S. Bell's concept of local causality}},  American
  Journal of Physics {\bf 79} (Dec., 2011) 1261--1275,
  [\href{http://xxx.lanl.gov/abs/0707.0401}{{\tt arXiv:0707.0401}}].

\bibitem{Preskill:1998}
J.~Preskill, {\it {Quantum Information and Computation}},  {Lecture notes for
  Physics 229 (Caltech)
  {\tt\url{http://www.theory.caltech.edu/people/preskill/ph229/}}}, 1998.

\bibitem{Bricmont:2014}
J.~Bricmont, {\it {What did Bell really prove?}},  Int. J. Quant. Found.
  (archive) {\tt\url{http://www.ijqf.org/archives/1303}}, 2014.

\bibitem{Bohm:1951}
D.~Bohm, {\em Quantum theory}.
\newblock Dover, Unknown, 1st~ed., 1989.

\bibitem{Gisin:2008}
D.~{Salart}, A.~{Baas}, J.~A.~W. {van Houwelingen}, N.~{Gisin}, and
  H.~{Zbinden}, {\it {Spacelike Separation in a Bell Test Assuming
  Gravitationally Induced Collapses}},  Physical Review Letters {\bf 100}
  (June, 2008) 220404, [\href{http://xxx.lanl.gov/abs/0803.2425}{{\tt
  arXiv:0803.2425}}].

\bibitem{Zeilinger:2013}
M.~{Giustina}, A.~{Mech}, S.~{Ramelow}, B.~{Wittmann}, J.~{Kofler}, J.~{Beyer},
  A.~{Lita}, B.~{Calkins}, T.~{Gerrits}, S.~W. {Nam}, R.~{Ursin}, and
  A.~{Zeilinger}, {\it {Bell violation using entangled photons without the
  fair-sampling assumption}},  \nat {\bf 497} (May, 2013) 227--230,
  [\href{http://xxx.lanl.gov/abs/1212.0533}{{\tt arXiv:1212.0533}}].

\bibitem{Hawking:1991nk}
S.~W. Hawking, {\it {The Chronology Protection Conjecture}},  Phys. Rev. {\bf
  D46} (1992) 603--611.

\bibitem{sep-qm-bohm}
S.~Goldstein, {\it Bohmian mechanics},  in {\em The Stanford Encyclopedia of
  Philosophy} (E.~N. Zalta, ed.).
\newblock spring 2013~ed., 2013.

\bibitem{Tumulka:2006}
R.~{Tumulka}, {\it {A Relativistic Version of the Ghirardi Rimini Weber
  Model}},  Journal of Statistical Physics {\bf 125} (Nov., 2006) 821--840,
  [\href{http://xxx.lanl.gov/abs/quant-ph/0406094}{{\tt quant-ph/0406094}}].

\bibitem{Bell:1990}
J.~S. Bell, {\it La nouvelle cuisine},  in {\em Between Science and Technology}
  (A.~Sarlemijn and P.~Kroes, eds.), pp.~97--115, North-Holland, 1990.

\bibitem{sep-time-travel-phys}
F.~Arntzenius and T.~Maudlin, {\it Time travel and modern physics},  in {\em
  The Stanford Encyclopedia of Philosophy} (E.~N. Zalta, ed.).
\newblock winter 2013~ed., 2013.

\bibitem{Everett:2012}
H.~Everett, J.~Barrett, and P.~Byrne, {\em The Everett Interpretation of
  Quantum Mechanics: Collected Works 1955-1980 with Commentary}.
\newblock high everett. Princeton University Press, 2012.

\bibitem{Albert:2009}
D.~Albert, {\em Quantum Mechanics and Experience}.
\newblock Harvard University Press, 2009.

\bibitem{Pegg:2008}
D.~T. Pegg, {\it Retrocausality and quantum mechanics},  Studies in History and
  Philosophy of Science Part B: Studies in History and Philosophy of Modern
  Physics {\bf 39} (2008), no.~4 830 -- 840.

\bibitem{Fine:1989}
A.~Fine, {\it Do correlations need to be explained?},  in {\em Philosophical
  Consequences of Quantum Theory: Reflections on Bell's theorem} (J.~T. Cushing
  and E.~McMullin, eds.), pp.~175--194.
\newblock University of Notre Dame Press, 1989.

\bibitem{Maccone:2015}
L.~Maccone. Personal Communication, 2015.

\bibitem{Durr:2013asa}
D.~Dürr, S.~Goldstein, T.~Norsen, W.~Struyve, and N.~Zanghì, {\it {Can
  Bohmian mechanics be made relativistic?}},  Proc.Roy.Soc.Lond. {\bf A470}
  (2013) 20130699, [\href{http://xxx.lanl.gov/abs/1307.1714}{{\tt
  arXiv:1307.1714}}].

\bibitem{Dickson:1998}
W.~Dickson, {\em Quantum chance and non-locality}.
\newblock Cambridge University Press, 1998.

\end{thebibliography}\endgroup

\appendix

\section{Relationship to Bell's original inequality}
\label{sec:Bell}

Here we describe how our informally derived inequality Eq.~\eqn{Bell-ineq} 
follows from the mathematical inequality derived by Bell.
%The EPR experiment and our notation for its results
%are shown Fig.~\ref{fig:EPR_polarizer}.
%
%\newcommand\prob[4]{\mbox{prob}(#1\!:#2,#3\!:#4)}
\newcommand\prob[4]{p(#2#4|#1#3)}
\newcommand\numb[4]{N(#1\!:#2,#3\!:#4)}
\newcommand\pdiff{p_{\rm diff}}
Let us define
\beq
\ba{l}
\prob{A}{+}{B}{-} \equiv \\[-1ex]
\quad\parbox{18em}{\begin{flushleft}
probability that, {\em given} that machines 1 and 2 have decided to deploy
filters A and B respectively, photon 1 passes through and photon 2
bounces off
\end{flushleft}}
\ea
\label{eq:notation} 
\eeq
and so on. Then Bell's original inequality is
\beq
\ba{lcl}
\prob A+B- + \prob B+C- + \prob C+A- &\leqslant& 1 \\[0.5ex]
\mbox{or, equivalently,}\\[0.5ex]
\prob A-B+ + \prob B-C+ + \prob C-A+ &\leqslant& 1 \ .
\ea
\label{Bell-orig}
\eeq
To derive our inequality \eqn{Bell-ineq} from Bell's original inequality,
start by rewriting \eqn{Bell-ineq}  as
\beq
\ba{rcl} 
\pdiff &\leqslant& 2/3 \\[1ex]
\mbox{where}\\[-1ex]
 \pdiff &=&
\mbox{p}\biggl(\bt{l}
   \mbox{when photon 1 and photon 2}\\[-0.5ex]
   \mbox{encounter different filters,}\\[-0.5ex]
   \mbox{they show different behavior} 
  \et 
  \biggr)\ ,  %\leqslant 2/3
\ea
\label{eq:pdiff}
\eeq
then using the notation \eqn{eq:notation} and defining
$p(AB)$ to be the
probability that machine 1 deploys the $A$ filter and machine $2$ deploys
the $B$ filter, and so on, we can rewrite
$\pdiff$ as a sum over all 
filter settings $F=(AB,BC,CA,BA,CB,AC)$ 
where the two detectors deploy different filters:
\beq
\pdiff = \frac{\sum_F p(F) \Bigl(p(+-|F)+p(-+|F)\Bigl)}{\sum_F p(F)} \ .
% \ba{rrl}
%        & \mbox{p}(AB)\bigl(\prob A+B- + \prob A-B+\bigr) \\[0.5ex]
%    +   & \mbox{p}(BA)\bigl(\prob B+A- + \prob B-A+\bigr) \\[0.5ex]
%    +   & \mbox{p}(BC)\bigl(\prob B+C- + \prob B-C+\bigr) \\[0.5ex]
%    +   & \mbox{p}(CB)\bigl(\prob C+B- + \prob C-B+\bigr) \\[0.5ex]
%    +   & \mbox{p}(CA)\bigl(\prob C+A- + \prob C-A+\bigr) \\[0.5ex]
%    +   & \mbox{p}(AC)\bigl(\prob A+C- + \prob A-C+\bigr)
% \ea
\label{Bell-baroque}
\eeq
In our experiment, each filter is deployed at random, so all six combinations
occur with equal probability, 
\beq
\pdiff = \frac{1}{6} \sum_F \Bigl(p(+-|F)+p(-+|F)\Bigl) \ .
\label{eq:pdiff-prob}
\eeq
The labeling of photons and measuring machines
as ``1'' and ``2'' is arbitrary, so with no loss of generality we
can treat the $BA$ filter deployment as being $AB$ with the numbering
of the photons and machines reversed, so $\prob A+B- = \prob B-A+$ and so on,
so \eqn{eq:pdiff-prob} can be written
\beq
\ba{rl}
\pdiff = \dsp\frac{1}{3} \Bigl( & 
  \prob A+B- + \prob B+C- + \prob C+A- \\
 + & \prob A-B+ + \prob B-C+ + \prob C-A+ \Bigr) \ .
\ea
\eeq
Using Bell's original inequality \eqn{Bell-orig} we recover our
inequality \eqn{eq:pdiff}.

\begin{figure}[tb]
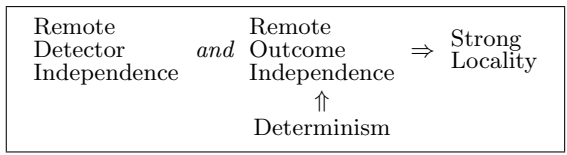

\fbox{
\raggedright
\begin{tabular}{ccccc}
  \bt{l} Remote\\[-0.5ex] Detector\\[-0.5ex] Independence\et
& {\em and}
& \bt{l} Remote\\[-0.5ex] Outcome\\[-0.5ex] Independence\et
& $\Rightarrow$
& {\bt{l}Strong\\[-0.5ex] Locality \et} \\
&& $\Uparrow$ \\
&& Determinism
\end{tabular}
}
\caption{Strong locality can be understood as saying that
the outcome at one detector is independent of both the settings
of the remote detector, and the outcome of the remote 
measurement \cite{Jarrett:1989}.
\label{fig:Jarrett}}
\end{figure}

\section{Remote Detector vs.~Remote Outcome independence}
\label{sec:remote}

As we saw in Sec.~\ref{sec:signal-locality}, there are two forms
of non-controllability that would allow us to keep 
the desirable principle of signal locality
in the face of the EPRB experiment's results.
Jarrett  \cite{Jarrett:1989}
pointed out one way to approach this. As illustrated in Fig.~\ref{fig:Jarrett},
strong locality can be written as a combination of two
requirements on the outcome of measurements at a given detector.
\ben
\item{\em Parameter independence},
which could be more informatively described as
{\em Remote Detector independence}:
the outcome of the measurement
of one photon is not affected by the detector setting for the
other photon, but may be affected by the measurement outcome
of the other photon. As we will describe
below, this form of locality is compatible with EPR
experiments and signal locality, 
but requires us to give up determinism.
\item{\em Remote Outcome independence}: the outcome of the measurement
of one photon is not affected by the outcome of the remote measurement,
but may be affected by the remote detector setting. This allows us
to keep determinism, but requires some limits on our ability to control 
the hidden variables
whose states determine measurement outcomes.
\een
%
%\cite{sep-qm-action-distance,sep-bell-theorem,Winsberg:2003}.
We can then understand EPR experiments as allowing us
keep no more than one of these requirements.\\[1ex]

The possibilities are:\\
(a) Parameter independence
is preserved, but Remote Outcome independence is violated. 
This corresponds to possibility (a) in Sec.~\ref{sec:signal-locality}.
We have to give up determinism because
Remote Outcome independence can be shown to follow from determinism
(Fig.~\ref{fig:Jarrett}).
Since the results of measurements are independent of the
controllable aspects of the remote experiment (its detector
settings) we preserve signal locality. The violation
of strong locality is achieved via a
superluminal influence that allows the 
(indeterministic and therefore uncontrollable) {\em outcome}
of one measurement to influence the other.
Parameter independence is a nice guarantee of signal
locality, but, as Bell pointed out (see the quote at the start of
Sec.~\ref{sec:signal-locality}),
it is unclear what is the fundamental basis for
this partition of phenomena into ``parameters'' and ``outcomes''.
Quantum Mechanics is an example of a theory that preserves
parameter independence while violating
Remote Outcome independence.\\[1ex]
(b) Remote Outcome independence is preserved, but
Parameter independence is violated. 
This corresponds to (b) in Sec.~\ref{sec:signal-locality}.
Allowing measurement results to be influenced by the settings of the remote
detector allows us to account for the EPRB results
while keeping  determinism, so there are
hidden variables whose state determines the outcome of measurements,
but to keep signal locality there must be
essential limits on our ability to
control the state of the hidden variables, and it is not
clear what general physical principle would ensure this.
Bohmian Mechanics is an example of a theory that,
{\it pace} Dickson \cite{Dickson:1998}, 
is usually said to violate Parameter independence while preserving
Remote Outcome independence \cite{Maudlin:2011}.

\end{document}